\documentclass[apj]{emulateapj}
\usepackage{rotate,amssymb}
\def\lsim{\ifmmode\stackrel{<}{_{\sim}}\else$\stackrel{<}{_{\sim}}$\fi}
\def\gsim{\ifmmode\stackrel{>}{_{\sim}}\else$\stackrel{<}{_{\sim}}$\fi}

\begin{document}

\journalinfo{The Astrophysical Journal, in press}
\submitted{The Astrophysical Journal, in press}

\title{High-Resolution Timing Observations of Spin-Powered Pulsars
 with the \emph{AGILE} Gamma-Ray Telescope\altaffilmark{1}}

\shorttitle{\emph{AGILE} Observations of Spin-Powered Pulsars}
\shortauthors{A.~Pellizzoni et al.}

%
\author{
A.~Pellizzoni,\altaffilmark{2}
M.~Pilia,\altaffilmark{2,3}
A.~Possenti,\altaffilmark{3}
F.~Fornari,\altaffilmark{2}
P.~Caraveo,\altaffilmark{2}
E.~Del Monte,\altaffilmark{4}
S.~Mereghetti,\altaffilmark{2}
M.~Tavani,\altaffilmark{4,5}
A.~Argan,\altaffilmark{4}
A.~Trois,\altaffilmark{4}
M.~Burgay,\altaffilmark{3}
A.~Chen,\altaffilmark{2,6}
I.~Cognard,\altaffilmark{7}
E.~Costa,\altaffilmark{4}
N.~D'Amico,\altaffilmark{3,8}
P.~Esposito,\altaffilmark{2,9,10}
Y.~Evangelista,\altaffilmark{4}
M.~Feroci,\altaffilmark{4}
F.~Fuschino,\altaffilmark{11}
A.~Giuliani,\altaffilmark{2}
J.~Halpern,\altaffilmark{12}
G.~Hobbs,\altaffilmark{13}
A.~Hotan,\altaffilmark{14}
S.~Johnston,\altaffilmark{13}
M.~Kramer,\altaffilmark{15}
F.~Longo,\altaffilmark{16}
R.~N.~Manchester,\altaffilmark{13}
M.~Marisaldi,\altaffilmark{11}
J.~Palfreyman,\altaffilmark{17}
P.~Weltevrede,\altaffilmark{13}
G.~Barbiellini,\altaffilmark{16}
F.~Boffelli,\altaffilmark{9,10}
A.~Bulgarelli,\altaffilmark{11}
P. W. Cattaneo,\altaffilmark{9}
V.~Cocco,\altaffilmark{4}
F.~D'Ammando,\altaffilmark{4,5}
G.~De Paris,\altaffilmark{4}
G.~Di Cocco,\altaffilmark{4}
I.~Donnarumma,\altaffilmark{4}
M.~Fiorini,\altaffilmark{2}
T.~Froysland,\altaffilmark{5,6}
M.~Galli,\altaffilmark{18}
F.~Gianotti,\altaffilmark{11}
A.~Harding,\altaffilmark{19}
C.~Labanti,\altaffilmark{11}
I.~Lapshov,\altaffilmark{4}
F.~Lazzarotto,\altaffilmark{4}
P.~Lipari,\altaffilmark{20}
F.~Mauri,\altaffilmark{9}
A.~Morselli,\altaffilmark{21}
L.~Pacciani,\altaffilmark{4}
F.~Perotti,\altaffilmark{2}
P.~Picozza,\altaffilmark{21}
M.~Prest,\altaffilmark{22}
G.~Pucella,\altaffilmark{4}
M.~Rapisarda,\altaffilmark{23}
A. Rappoldi,\altaffilmark{9}
P.~Soffitta,\altaffilmark{4}
M.~Trifoglio,\altaffilmark{11}
E.~Vallazza,\altaffilmark{16}
S.~Vercellone,\altaffilmark{2}
V.~Vittorini,\altaffilmark{5}
A.~Zambra,\altaffilmark{2}
D.~Zanello,\altaffilmark{20}
C.~Pittori,\altaffilmark{24}
F.~Verrecchia,\altaffilmark{24}
B.~Preger,\altaffilmark{24}
P.~Santolamazza,\altaffilmark{24}
P.~Giommi,\altaffilmark{24}
and
L.~Salotti\altaffilmark{25}
}

\altaffiltext{1}{Based on
 observations obtained with \emph{AGILE}, an ASI (Italian Space
 Agency) science mission with instruments and contributions directly
 funded by ASI.}

\altaffiltext{2}{INAF/IASF--Milano, via E.~Bassini 15, I-20133 Milano,
Italy} 
\altaffiltext{3}{INAF--Osservatorio Astronomico di Cagliari, 
localit\'a Poggio dei Pini, strada 54, I-09012 Capoterra, Italy}
\altaffiltext{4}{INAF/IASF--Roma,
via del Fosso del Cavaliere 100, I-00133 Roma, Italy}
\altaffiltext{5}{Dipartimento di Fisica, Universit\`a ``Tor Vergata'', via della Ricerca 
Scientifica 1, I-00133 Roma, Italy} 
\altaffiltext{6}{CIFS--Torino, viale Settimio Severo 3,
I-10133, Torino, Italy} 
\altaffiltext{7}{Laboratoire de Physique et Chimie de l'Environnement, 
CNRS, F-45071 Orleans, France}
\altaffiltext{8}{Dipartimento di Fisica, Universit\`a di Cagliari,
Cittadella Universitaria, I-09042 Monserrato, Italy}
\altaffiltext{9}{INFN--Pavia, via Bassi 6, I-27100 Pavia, Italy}
\altaffiltext{10}{Dipartimento di Fisica Nucleare e Teorica, 
Universit\`a di Pavia, via A.~Bassi 6, Pavia, I-27100, Italy}
\altaffiltext{11}{INAF/IASF--Bologna, via
Gobetti 101, I-40129 Bologna, Italy} 
\altaffiltext{12}{Columbia Astrophysics Laboratory, Columbia University, 
New York, NY 10027 }
\altaffiltext{13}{Australia Telescope National Facility, CSIRO, 
P.O.~Box~76, Epping NSW~1710, Australia}
\altaffiltext{14}{Curtin University of Technology, 78 Murray Street,
Perth, WA 6000, Australia}
\altaffiltext{15}{University of Manchester, Jodrell Bank Observatory, 
Macclesfield, Cheshire SK11 9DL, UK}
\altaffiltext{16}{Dipartimento di Fisica, Universit\`a di Trieste and INFN--Trieste, via Valerio 2, I-34127 Trieste, 
Italy}
\altaffiltext{17}{School of Mathematics and Physics, University of 
Tasmania, Hobart, TAS 7001, Australia}
\altaffiltext{18}{ENEA--Bologna, via Biancafarina 2521, 
I-40059 Medicina (BO), Italy} 
\altaffiltext{19}{Astrophysics Science Division, NASA/Goddard Space 
Flight Center, Greenbelt, MD 20771}
\altaffiltext{20}{INFN--Roma ``La Sapienza'', piazzale A. Moro 2, 
I-00185 Roma, Italy}
\altaffiltext{21}{INFN--Roma ``Tor Vergata'', via della Ricerca 
Scientifica 1, I-00133 Roma, Italy}
\altaffiltext{22}{Dipartimento di Fisica, Universit\`a dell'Insubria, via Valleggio 11, 
I-22100 Como, Italy}
\altaffiltext{23}{ENEA--Roma, via E. Fermi 45, I-00044 Frascati (Roma), 
Italy}
\altaffiltext{24}{ASI--ASDC, via G. Galilei, I-00044 Frascati (Roma), 
Italy}
\altaffiltext{25}{ASI, viale Liegi 26, I-00198 Roma, Italy}

\email{alberto@iasf-milano.inaf.it}

\begin{abstract}

\emph{AGILE} is a small gamma-ray astronomy satellite mission of the
Italian Space Agency dedicated to high-energy astrophysics launched in
2007 April. Its $\sim$1 $\mu$s absolute time tagging capability
coupled with a good sensitivity in the 30 MeV--30 GeV range, with
simultaneous X-ray monitoring in the 18--60 keV band, makes it
perfectly suited for the study of gamma-ray pulsars following up on
the \emph{CGRO}/EGRET heritage.  In this paper we present the first
\emph{AGILE} timing results on the known gamma-ray pulsars Vela, Crab,
Geminga and B\,1706--44. The data were collected from 2007 July to 2008
April, exploiting the mission Science Verification Phase, the
Instrument Timing Calibration and the early Observing Pointing
Program.  Thanks to its large field of view, \emph{AGILE} collected a
large number of gamma-ray photons from these pulsars ($\sim$10,000
pulsed counts for Vela) in only few months of observations.  The
coupling of \emph{AGILE} timing capabilities, simultaneous radio/X-ray
monitoring and new tools aimed at precise photon phasing, exploiting
also timing noise correction, unveiled new interesting features at
sub-millisecond level in the
pulsars' high-energy light-curves.
\end{abstract}

\section{Introduction}

Among the $\sim$1,800 known rotation-powered pulsars, observed mainly
in the radio band, seven objects have been identified as gamma-ray
emitters, namely Vela (B\,0833--45), Crab (B\,0531+21), Geminga
(J\,0633+1746), B\,1706--44, B\,1509--58, B\,1055--52 and B\,1951+32
\citep[][]{thompson04}. In addition, B\,1046--58 \citep{klm00},
B\,0656+14 \citep{ramanamurthy96} and J\,0218+4232
\citep{kuiper00,kuiper02} were reported with lower confidence
(probability of periodic signal occurring by chance in gamma-rays of
$\sim$10$^{-4}$). In spite of the paucity of pulsar
identifications, gamma-ray observations are a valuable tool for
studying particle acceleration sites and emission mechanisms in the
magnetospheres of spin-powered pulsars.

So far, spin-powered pulsars were the only class of Galactic sources
firmly identified by \emph{CGRO}/EGRET and presumably some of the
unidentified gamma-ray sources will turn out to be
associated to young and energetic radio pulsars discovered in recent
radio surveys \citep{manchester01,kramer03}. In fact,
several unidentified gamma-ray sources
have characteristics similar to those of the known gamma-ray pulsars
(hard spectrum with high-energy cutoff, no variability, possible X-ray
counterparts with thermal/non-thermal component, no prominent optical
counterpart), but they lack a radio counterpart as well as a supernova
remnant and/or pulsar wind nebula association. Radio quiet,
Geminga-like objects have been invoked by several authors
\citep{romani95,yadigaroglu95,harding07} but without evidence of
pulsation in gamma-rays, no identification has been confirmed.

Apart from the Crab and B\,1509--58, whose luminosities peak in the
100 keV and about 30 MeV range respectively, the energy flux of the remaining gamma-ray
pulsars is dominated by the emission above 10 MeV with a spectral
break in the GeV range.

\emph{AGILE} (Astro-rivelatore Gamma ad Immagini LEggero) is a small
scientific mission of the Italian Space Agency dedicated to
high-energy astrophysics \citep[][Tavani et al. 2008, in preparation]{tavani06} launched on 2007 April 23. Its
sensitivity in the 30 MeV--30 GeV range, with simultaneous X-ray
imaging in the 18--60 keV band, makes it perfectly suited for the
study of gamma-ray pulsars.  Despite its small dimensions and weight
($\sim$100 kg), the new silicon detector technology employed for the
\emph{AGILE} instruments yields overall performances as good as, or
better than, that of previous bigger instruments. High-energy photons
are converted into e$^{+}$/e$^{-}$ pairs in the Gamma-Ray Imaging
Detector (GRID) a Silicon-Tungsten tracker
\citep{prest03,barbiellini01}, allowing for an efficient photon
collection, with an effective area of $\sim$500 cm$^{2}$, and for an
accurate arrival direction reconstruction ($\sim$0.5$^{\circ}$ at 1
GeV) over a very large field of view, covering about 1/5 of the sky in
a single pointing. The Cesium-Iodide mini-calorimeter
\citep{labanti06} is used in conjunction with the tracker for photon
energy reconstruction while supporting the anti-coincidence shield in
the particle background rejection task \citep{perotti06}.
The \emph{AGILE}/GRID
is characterized by the smallest dead-time ever obtained for gamma-ray
detection (typically 200 $\mu$s) and time tagging with uncertainty
near $\sim$1 $\mu$s.
The SuperAGILE hard X-ray monitor is positioned on top of
the GRID. SuperAGILE
is a coded aperture instrument operating in the 18--60 keV energy band 
with about 15 mCrab sensitivity in one
day integration, 6 arcmin angular resolution
and $\sim$1 sr field of view \citep{feroci07,costa01}. 

In this work we analyze all available \emph{AGILE}/GRID data suitable
for timing analysis collected up to 2008 April 10 for the four known
gamma-ray pulsars included in the \emph{AGILE} Team source
list:\footnote{See http://agile.asdc.asi.it for details about
\emph{AGILE} Data Policy and Target List.} Vela, Crab, Geminga and
B\,1706--44. The other two EGRET pulsars, B\,1055--52 and B\,1951+32,
are part of the \emph{AGILE} Guest Observer program. As expected, only the Crab
pulsar has been detected by SuperAGILE and the X-ray data have been
used to cross-check and test \emph{AGILE} timing performances.

The \emph{AGILE} observations are presented in \S~\ref{datareduction}, 
as well as the criteria for photons selections. The observations and the
timing analysis from the parallel radio and X-ray observations of the
four targets are described in \S~\ref{sec:radio}. The procedures
for the timing analysis of gamma-ray data are introduced in 
\S~\ref{timinganalysis} and the results of their application are
reported in \S~\ref{section-tctr}, where timing calibration tests
are also dealt with. Discussion of the scientific results and
conclusions are the subjects of \S\S~\ref{discuss} and \ref{conclusions}, 
respectively.

\section{\emph{AGILE} Observations and Data Reduction}
\label{datareduction}

The \emph{AGILE} spacecraft was placed in a Low Earth Orbit (LEO) at
$\sim$535 km mean altitude with inclination $\sim$2.5\degr. Therefore,
Earth occultation strongly affects exposure along the orbital plane,
as well as high particle background rate during South Atlantic Anomaly
transits. However, the exposure efficiency is $>$50\% for most
\emph{AGILE} revolutions.  \emph{AGILE} pointings consist of long
exposures (typically lasting 10--30 days) slightly drifting
($\lesssim$1\degr/day) with respect to the starting pointing direction in order
to match solar-panels illumination constraints. The relatively uniform
values of the effective area and point spread function within
$\sim$40\degr\ from the center of the field of view of the GRID, allow for
one-month pointings without significant vignetting in the exposure of
the target region.

\begin{figure*}  
\centering
\resizebox{.8\hsize}{!}{\includegraphics[angle=00]{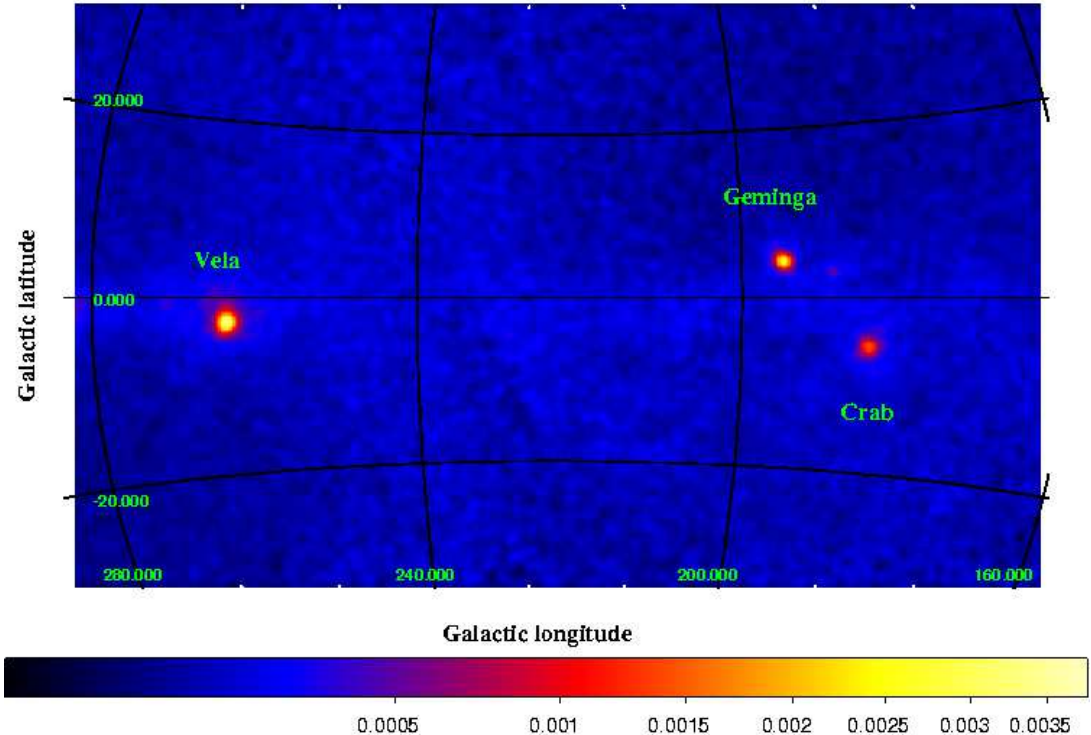}}
\caption{\label{pulsarmap} Gaussian-smoothed \emph{AGILE} intensity
map ($\sim$120\degr\ $\times$ 60\degr, units: ph cm$^{-2}$ s$^{-1}$ sr$^{-1}$, E$>$100 MeV) in Galactic coordinates
integrated over the whole observing period (2007 July 13--2008 April
10) and centerd at $l=223\degr$, $b=0\degr$. The \emph{AGILE} field of
view (radius $\sim$60\degr) can embrace in a single pointing Vela ($l=263.6\degr$, $b=-2.8\degr$),
Geminga ($l=195.1\degr$, $b=4.3\degr$) and Crab ($l=184.6\degr$, $b=-5.8\degr$) as well as diffuse emission from the Galactic Disk.} 
\end{figure*}

Pulsar data were collected during the mission Science Verification
Phase (SVP, 2007 July--November) and early pointings\footnote{1. Cygnus
Field 1 ($l=89\degr$, $b=9.9\degr$), 2. Virgo Field ($l=264\degr$,
$b=56.5\degr$), 3. Vela Field ($l=283\degr$, $b=-6.8\degr$), 4. South
Gal.  Pole ($l=240\degr$, $b=-50\degr$), 5. Musca Field ($l=303\degr$,
$b=-9\degr$), 6. Gal. Center ($l=332\degr$, $b=0\degr$),
7. Anti-Center ($l=193\degr$, $b=8.1\degr$).} (2007 December--2008 April) 
of the AO\,1 Observing Program. It is worth noting that a single
\emph{AGILE} pointing on the Galactic Plane embraces about one-third
of it, allowing for simultaneous multiple source targeting
(e.g. the Vela and Anti-Center regions in the same field of view with
Crab, Geminga and Vela being observed at once; Figure \ref{pulsarmap}).

The \emph{AGILE} Commissioning and Science Verification Phases
lasted about seven months from 2007 April 23 to November 30 including
also Instrument Time Calibration. On 2007 December 1 baseline nominal
observations and a pointing plan started together with the Guest
Observer program AO\,1. Timing observations suitable for pulsed signal
analysis of the Vela pulsar started in mid 2007 July (at orbit 1,146)
after engineering tests on the payload.

\begin{figure} 
\resizebox{\hsize}{!}{\includegraphics[angle=00]{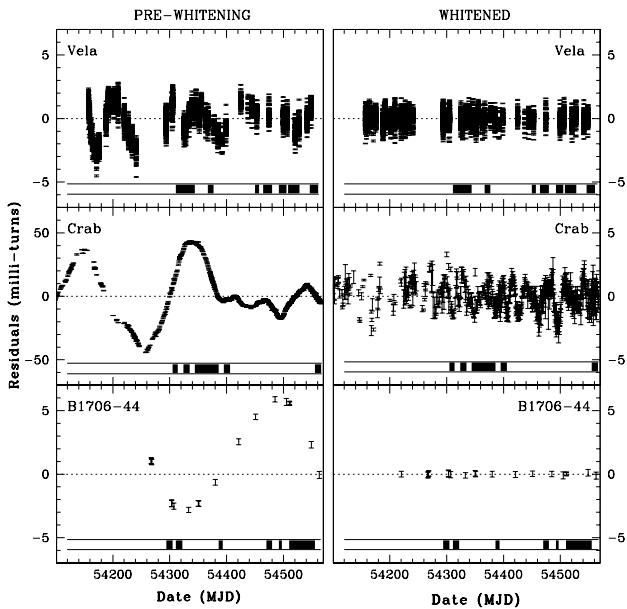}}
\caption{\label{fig:residual} Post-fit timing residuals (in
milli-turns) as a function of the Modified Julian Day
resulting from the observation at 1.4~GHz of the three radio pulsars
which are discussed in this paper: from the top Vela, Crab and
PSR~B1706$-$44. The panels in the left (right) column report the
residuals of the best available timing solutions obtained over the
data span not including (including) the correction of the timing noise
via the use of the $\Delta R$ term (see \S~\ref{timinganalysis}).
Note that for the Crab pulsar the scale on the vertical axis of the
panel in the right column is amplified by a factor 10 with respect to
that in the left panel.  The time intervals corresponding to the
useful \emph{AGILE} pointings for each target are also given as the
black sections of the bar at the bottom of each panel.}
\end{figure}

The Vela region was observed (with optimal exposure efficiency) for $\sim$40 
days during the SVP and again for $\sim$30 days in AO\,1 pointing number 3 
(2008 January 8--February 1). PSR B\,1706--44 was within the Vela and Galactic 
Center pointings for $\sim$30 days during the SVP and for $\sim$45 days
during AO\,1 pointings numbers 5 and 6 (2008 February 14--March 30). The
Anti-center region (including Crab and Geminga) was observed for $\sim$40
days, mostly in 2007 September, and in 2008 April (AO\,1 pointing
number 7) with the addition of other sparse short Crab pointings for
SuperAGILE calibration purposes during the SVP (see Table \ref{counts}
and Figure \ref{fig:residual} for details about targets coverage).
\begin{deluxetable*}{clccccc}  
  \tablecolumns{1}
\tablewidth{0pc}
 \tablecaption{\label{counts} Observation parameters of the relevant
data subsets (grouped in uniterrupted target observations) for each
analyzed pulsar considering for comparison all G+L class events (see
text) with $E>100$ MeV observed within 60\degr\ from the center of the
field of view and extracted within 5\degr\ from the pulsar position.}
\tablehead{
\colhead{PSR} & \colhead{ObsID\tablenotemark{a}} & \colhead{T$_{FIRST}$} & \colhead{T$_{LAST}$} & \colhead{$\langle\theta\rangle$\tablenotemark{b}} & \colhead{Total Counts\tablenotemark{c}} & \colhead{Exposure\tablenotemark{d} } \\
\colhead{ } & \colhead{ } & \colhead{[MJD]} & \colhead{[MJD]} & \colhead{[deg]} & \colhead{ } & \colhead{[10$^8$ cm$^2$ s]} }

\startdata

Vela & SVP 1 & 54,294.552 & 54,305.510 & 26.2 & 6440 & 1.87 \\
Vela & SVP 2-3-4 & 54,311.512 & 54,344.569 & 40.9 & 12651 & 3.26 \\
Vela & SVP 5 & 54,358.525 & 54,359.521 & 41.4 & 391  & 0.11 \\
Vela & SVP 6 & 54,367.525 & 54,377.521 & 48.9 & 1355 & 0.26 \\
Vela & SVP 8 & 54,395.554 & 54,406.501 & 55.3 & 113  & 0.03 \\
Vela & AO\,1 2 & 54,450.540 & 54,457.157 & 56.2 & 564  & 0.01 \\
Vela & AO\,1 2-3 & 54,464.784 & 54,480.966 & 22.1 & 4726 & 3.13 \\
Vela & AO\,1 3-4 & 54,492.948 & 54,505.377 & 43.2 & 6293 & 1.16 \\
Vela & AO\,1 4-5-6 & 54,508.528 & 54,528.537 & 43.6 & 9747 & 2.22 \\
Vela & AO\,1 6 & 54,546.095 & 54,561.427 & 41.7 & 703 & 0.17 \\
Vela & total\tablenotemark{e} & 54,294.552 & 54,561.427 & 36.1 & 42983 & 12.22 \\
Vela & gamma\tablenotemark{f} & 54,294.552 & 54,561.427 & 36.1 & 6140 & 5.28 \\

\hline

Geminga  & SVP 2 & 54,308.871 & 54,314.507 & 55.0 & 187 & 0.06  \\
Geminga  & SVP 3 & 54,324.514 & 54,335.508 & 41.3 & 781 & 0.24 \\
Geminga  & SVP 4-5-6-7 & 54,344.518 & 54,351.235 & 22.2 & 17823 & 6.19 \\
Geminga  & SVP 8 & 54,395.520 & 54,406.504 & 37.6 & 1926 & 0.48 \\
Geminga  & AO\,1 6 & 54,528.531 & 54,529.270 & 50.7 & 38   & 0.01 \\
Geminga  & AO\,1 6 & 54,546.093 & 54,549.428 & 39.0 & 132 & 0.03 \\
Geminga  & AO\,1 6-7 & 54,555.514 & 54,566.5 & 11.7 & 4860 & 2.03 \\
Geminga  & total\tablenotemark{e} & 54,308.871 & 54,566.5  & 20.2 & 25962 & 9.04 \\
Geminga  & gamma\tablenotemark{f} & 54,308.871 & 54,566.5 & 20.2 & 3874 & 4.54 \\

\hline

Crab & SVP 1-2 & 54,305.509 & 54,314.50 & 50.2 & 2404 & 0.58 \\
Crab & SVP 3 & 54,324.512 & 54,335.508 & 28.2 & 761 &.0.32 \\
Crab & SVP 4-5-6-7 & 54,344.518 & 54,386.502 & 23.9 & 20499 & 6.41 \\
Crab & SVP 8 & 54,395.518 & 54,406.506 & 45.4 & 1537 & 0.32 \\
Crab & AO\,1 3-4 & 54,494.447 & 54,505.390 & 55.8 & 350  & 0.05 \\
Crab & AO\,1 4 & 54,508.506 & 54,510.474 & 54.7 & 359  & 0.06 \\
Crab & AO\,1 6 & 54,528.531 & 54,529.140 & 46.9 & 100  & 0.02 \\
Crab & AO\,1 6 & 54,546.091 & 54,549.423 & 41.5 & 176 & 0.03 \\
Crab & AO\,1 6-7 & 54,555.516 & 54,566.5 & 21.8 & 4973 & 1.89 \\
Crab & total\tablenotemark{e} & 54,305.509 & 54,566.5 & 26.1 & 31159 & 9.68 \\
Crab & gamma\tablenotemark{f} & 54,305.509 & 54,566.5 & 26.1 & 4062  & 4.11 \\

\hline

B1706--44  & SVP 1 & 54,294.552 & 54,305.503 & 50.9 & 4641 & 0.91 \\
B1706--44  & SVP 2-3-4 & 54,311.531 & 54,322.891 & 34.5 & 15657 & 4.39 \\
B1706--44  & SVP 7-8 & 54,386.524 & 54,393.752 & 41.5 & 5521 & 1.28 \\
B1706--44  & AO\,1 2-3 & 54,470.292 & 54,480.965 & 53.7 & 2169 & 1.46 \\
B1706--44  & AO\,1 3 & 54,492.948 & 54,497.509 & 44.9 & 1507 & 0.20 \\
B1706--44  & AO\,1 5-6 & 54,510.482 & 54,555.509 & 22.4 & 24219  & 7.36 \\
B1706--44  & total\tablenotemark{e} & 54,294.552 & 54,555.509 & 28.7 & 53714 & 15.6 \\
B1706--44  & gamma\tablenotemark{f} & 54,294.552 & 54,555.509 & 28.7 & 8463 & \phantom{6.56}6.56\enddata
\tablenotetext{a}{Observation ID: SVP=Science Verification Phase
(grouped in subset of 200 orbit each starting from 1,146),
AO\,1=Scientific Observations Program pointings (see \emph{AGILE}
Mission Announcement of Opportunity Cycle-1:
http://agile.asdc.asi.it/).}
\tablenotetext{b}{Mean off-axis angle.}
\tablenotetext{c}{Source photons + diffuse emission photons + particle
background.}
\tablenotetext{d}{Good observing time after dead-time and occultation
corrections.}
\tablenotetext{e}{Total G+L class events
with $E>100$ MeV, 5\degr\ max from pulsar position, 60\degr\ max
from field of view center.}
\tablenotetext{f}{High-confidence photon events (G class only) with $E>100$
MeV, 5\degr\ max from pulsar position, 60\degr\ max from field of view
center.}
\end{deluxetable*}

GRID data of the relevant observing periods were grouped in 20 subsets
of 200 orbits each (corresponding to $\sim$15 days of observation)
starting from orbit 1,146 (54,294 MJD).  Data screening, particle
background filtering and event direction and energy reconstruction
were performed by the \emph{AGILE} Standard Analysis Pipeline
(BUILD-15) for each subset with an exposure $>$$10^6$ cm$^2$ s at
$E>100$ MeV. Observations affected by coarse pointing, non-nominal
settings or intense particle background (e.g. orbital passages in
South Atlantic Anomaly) and albedo events from the Earth's limb were 
excluded from the processing. 

A specific optimization on the events extraction parameters is
performed for each target in order to maximize the signal to noise
ratio for a pulsed signal. The optimal event extraction radius around
pulsar positions varies as a function of photon energy (and
then it is related to pulsar spectra) according to the Point Spread
Function.  However, for $E>100$ MeV broad band analysis, a fixed
extraction radius of $\sim$5\degr\ (a value slightly higher than the
Point Spread Function 68\% containment radius) produces comparable
results with respect to energy-dependent extraction.

Quality flags define different GRID event classes. The G event class
includes events identified with good confidence as photons. Such
selection criteria correspond to an effective area of $\sim$250 cm$^2$
above 100 MeV (for sources within 30 deg from the center of the
field-of-view). The L event class includes events typically affected by
an order of magnitude higher particle contamination than G, but
yielding an effective area of $\sim$500 cm$^2$ at $E>100$ MeV, if
grouped with the G class. We performed our timing analysis looking for
pulsed signals using both G class events and the combination of G+L
events.  In general, the signal-to-noise ratio of the pulsed signal is
maximized using photons collected within $\sim$40\degr\ from the
center of GRID field of view and selecting the event class G.  For
very strong sources, such as Vela, or sources located in low
background regions, it is possible to include also photons in the
40--60\degr\ off-axis range and belonging to the G+L event class
typically 
improving the count statistics by up to a factor of three (obviously 
implying also a much higher background),
without affecting or even improving the detection significance.
For each pulsar, Table \ref{counts} summarizes the exposure parameters
of the relevant data subsets grouped in uninterrupted target
observations. 
For simplicity and comparison, we reported in Table \ref{counts} and Table \ref{counts2}
observation and detection parameters obtained with the same extraction
criteria and energy range.
For all observations with the targets within 60\degr\
from the pointing direction, we list the average angular distance
$\langle\theta\rangle$ of the source position from the pointing direction, the
sum of all G+L events with $E>100$ MeV whose direction is within
5\degr\ from the pulsar position and the G only events (``gamma
photons''), selected with the same criteria. The choice to include all
G+L events yields a ``dirty'' data set: for the case of Vela,
$\sim$10\% of the total counts are ascribable to the Galactic Plane,
and $>$50\% are particle background ($\sim$10,000 source counts).
The contamination is reduced in
the ``gamma'' entry, characterized by $\sim$20\% diffuse emission and
$\sim$5\% particle background ($\sim$5,000 source counts).
 The corresponding exposure for each
data subset was calculated with the GRID scientific analysis task
AG\_ExpmapGen according with the above parameters. Summing
up the photon numbers, we see that the overall photon statistics
accumulated (accounting for background contamination) is comparable to
that collected by EGRET for the same four pulsars.
\begin{deluxetable*}{cccccrcc}   
  \tablecolumns{1}
\tablewidth{0pc}
 \tablecaption{\label{counts2} The photon harvest from the \emph{AGILE}
 observations of gamma-ray pulsars.}
\tablehead{
\colhead{PSR} & \colhead{P} & \colhead{$\dot{E}$} & \colhead{$d$} & \colhead{Pulsed Counts\tablenotemark{a}} & \colhead{$\chi^{2}_{r}$ (d.o.f)} & \colhead{Exposure\tablenotemark{b}} & \colhead{Pulsed Flux\tablenotemark{c}} \\
\colhead{ } & \colhead{(ms)} &\colhead{(erg s$^{-1}$)} &\colhead{(kpc)} &\colhead{ } & \colhead{ } & \colhead{(10$^8$ cm$^2$ s)} & \colhead{10$^{-8}$ ph. cm$^{-2}$ s$^{-1}$}}

\startdata


Vela & 89.3  & $6.92\times10^{36}$ & 0.29 & $9,170\pm580$	& 225.51 (9) & 12.22 & $940\pm60$\\
Geminga & 237.1 & $3.25\times10^{34}$ & 0.16 & $2,200\pm480$ & 10.44 (9) & 9.04 & $300\pm70$\\
Crab	 & 33.1 & $4.61\times10^{38}$ & 2.00 & $2,120\pm530$	& 10.71 (9)  & 9.68 & $270\pm70$\\
B1706--44& 102.5 & $3.41\times10^{36}$ & 1.82 & $2,370\pm720$& 9.11 (9)   & 15.6 & \phantom{10}$190\pm60$\enddata
\tablecomments{See text for details on data reduction and timing analysis.}
\tablenotetext{a}{Pulsed counts (G+L event class) with $E>100$ MeV,
 5\degr\ max from pulsar position, 60\degr\ max from field of view
 center, 10 bins.}
\tablenotetext{b}{Good observing time after dead-time and occultation
corrections.}
\tablenotetext{c}{Calculated with the expression $C_P f/E$, where
$C_p$=pulsed counts, $E$=exposure, $f$=factor accounting for source
counts at angular distance $>$5\degr\ from source position according
to the point spread function ($f\sim1.25$).}
\end{deluxetable*}

A maximum likelihood analysis (ALIKE task) on the \emph{AGILE}
data from the sky areas containing the four gamma-ray pulsars yielded
source positions, fluxes and spectra in good agreement with those
reported both in the EGRET catalogue for the corresponding 3EG sources
\citep{hartman99} and in the revised catalogue by \citet{casandjian08}. 
As an example, Figure \ref{pulsarmap} reports an \emph{AGILE} intensity 
map displaying Vela, Geminga and Crab.  Details on gamma-ray imaging and 
spectra of the four sources will be reported in future papers 
when count statistics significantly higher than that currently available
are collected and in-flight Calibration files are finalized.

In this paper we focus on timing analysis. However, it is worth
noticing that pulsed counts provide a gamma-ray pulsar flux
estimate independent from the likelihood analysis, as described in 
\S~\ref{section-tctr}.

\section{Radio/X-rays observations and timing} 
\label{sec:radio}

In order to perform \emph{AGILE} timing calibration through accurate
folding and phasing, as described in \S\S~\ref{timinganalysis}
and \ref{section-tctr}, pulsar timing solutions valid for the epoch of
\emph{AGILE} observations were required. Thus, a dedicated pulsar
monitoring campaign (that will continue during the whole \emph{AGILE}
mission) was undertaken, using two telescopes (namely
Jodrell Bank and Nan\c cay) of the European Pulsar Timing Array
(EPTA), as well as 
the Parkes radio telescope of the Commonwealth Scientific and Industrial Research Organisation (CSIRO) Australia Telescope
National Facility (ATNF) and the 26m Mt Pleasant radio telescope operated by the
University of Tasmania.

In particular, the observations of the Vela Pulsar have been secured
by the Mt.~Pleasant Radio Observatory. Data
were collected at a central frequency of 1.4~GHz. Given the pulsar
brightness, it is possible to extract a pulse Time of Arrival (ToA)
about every 10 seconds, so that a total of 4,098 ToAs have been
obtained for the time interval between 2007 February 26 (MJD 54,157) and
2008 March 23 (MJD 54,548), encompassing the whole time span of the
\emph{AGILE} observations. During this time interval the pulsar
experienced a small glitch (fractional frequency increment
$\Delta\nu_g / \nu \simeq 1.3 \times 10^{-9}$), which presumably
happened around 2007 August 1 (with an uncertainty of $\pm$3 days
due to the lack of observations around this period) (see \S~\ref{velaglitch}).   Ephemeris for the pre- and post-glitch time
intervals were then separately calculated and applied to the gamma-ray
folding procedure.

The observations of the Crab Pulsar have been provided by the
telescopes of Jodrell Bank and Nan\c cay.  At Jodrell Bank the Crab
Pulsar has been observed daily from 2006 December 09 (MJD 54,078) to 
2007 October 06 (MJD 54,379) and again from 2008 February 06 
(MJD 54,502) to 2008 April 10 (MJD 54,566) which result in 334 ToAs. 
The observations were mainly performed at 1.4~GHz with the 76 m Lovell 
telescope, with some data also taken with 
the 12-m telescope at a central frequency of 600~MHz.  The Nan\c cay
Radio Telescope (NRT) is a transit telescope with the equivalent area of a 93 m dish and observed at a
central frequency of 1.4~GHz over the time interval between 2006 December 05
(MJD 54,074) and 2007 September 07 (MJD 54,350), producing 64
ToAs. A timing solution was produced by joining the ToAs from the two
telescopes and accounting for the phase shift that naturally ensues from
data sets coming from different telescopes.

The observations of pulsar B\,1706$-$44 have been performed at the
64 m telescope at Parkes, in Australia, at the mean observing frequency
of 1.4~GHz. They produced a total of 20 ToAs over the interval from 2007 April
30 (MJD 54,220) to 2008 Apr 6 (MJD 54,562), thus covering the
whole \emph{AGILE} observing time span for this target.

The timing of all pulsars is performed using the TEMPO2 software
\citep{hobbs06,edwards06}.  It first converts the topocentric
TOAs to solar-system barycentric TOAs at 
infinite frequency\footnote{Dispersion measure is obtained as part of timing
solution for Crab (DM=56.76(1)), Vela (DM=68.15(2)) and from \citet{johnston95}
for B\,1706$-$44 (DM=75.69(5)).}
 (using the Jet Propulsion Laboratory [JPL] DE405 solar-system
ephemeris\footnote{See ftp://ssd.jpl.nasa.gov/pub/eph/export/DE405/de405.iom/.})
and then performs a multi-parameter least-square fit to determine the
pulsar parameters. The differences between the observed barycentric
ToAs and those estimated by the adopted timing model are represented
by the so-called $residuals$. The procedure is iterative and improves
with longer timescales of observations. An important feature developed
by TEMPO2 is the possibilty of accounting for the timing noise in the
fitting procedure (see \S~\ref{timinganalysis}). This is
particularly useful in timing the young pulsars: in fact most of them
suffer of quasi-random fluctuations (typically characterized by a very
red noise spectrum) in the rotational parameters, whose origin is
still debated. TEMPO2 corrects the effects of the timing noise on the
residuals by modeling its behaviour as a sum of harmonically
correlated sinusoidal waves (see Equation [\ref{phasing2}] in
\S~\ref{timinganalysis}) that is subtracted from the residuals.

The left panels (labeled as {\it pre-whitening}) in
Figure~\ref{fig:residual} reports the timing residuals obtained over the
data span of the radio observations without the correction for the
timing noise, whereas the right panels (labeled as {\it whitened})
show the residuals after the application of the correction.  The
comparison between the panels in the two columns shows that this
procedure has been very effective in removing the timing noise in
Vela, Crab and PSR~B1706$-$44. The impact of this whitening procedure
on the gamma-ray data analysis is discussed in \S~\ref{section-tctr}.

Geminga is a radio-quiet pulsar whose ephemeris can be obtained from
X-ray data.  Following the demise of \emph{CGRO}, Geminga was
regularly observed with \emph{XMM-Newton} in order to maintain its
ephemeris for use in analyzing observations at other wavelengths.  
We analyzed all eight observations of Geminga (1E\,0630+178,
PSR\,J\,0633+1746) taken with the X-ray 
(0.1--15 keV) cameras of \emph{XMM-Newton} \citep{jansen01} between
MJD 52,368 and MJD 54,534 (2002 April - 2008 March), with exposures
in the range 20-100 ks.


The data were processed using version 7.1.0 of the \emph{XMM-Newton} 
Science Analysis Software (SAS) and the calibration files released in 2007 August.
For the timing analysis we could only use the pn data (operated in
Small Window mode: time resolution 6 ms, imaging on a
$4\arcmin\times4\arcmin$ field), owing to the inadequate time
resolution of the MOS data. We selected only single and double photon
events (patterns 0--4) and applied standard data screening
criteria. Photons arrival times were converted to the Solar system
barycenter using the SAS task barycen. To extract the source photons
we selected a circle of 30$\arcsec$ radius, containing about 85\% of
the source counts. Using standard folding and phase-fitting
techniques, the source pulsations were clearly detected in all the
observations. We derived for Geminga the long-term spin parameters
valid for the epoch range 52,369-54,534 MJD.
Note that the absolute accuracy
of the \emph{XMM-Newton} clock is better than 600 $\mu$s
\citep{kirsch04timing}.

The Crab Pulsar is embedded in the Crab Nebula and represents about
10\% of its flux in the hard X-ray band. Being a bright source with a
relatively soft high-energy spectrum, the Crab pulsar is easily
detected by SuperAGILE in less than one \emph{AGILE} orbital
revolution. Since the on-board time reconstruction of SuperAGILE
\citep{feroci07}  is different from that of the GRID, as a cross-check and test
of the SuperAGILE timing performances we processed the X-ray data of
an on-axis observation of $\sim$0.7 days (54,360.7--54,361.4 MJD),
corresponding to an effective exposure of $\sim$41 ks. In the analysed
data the passages of \emph{AGILE} in the South Atlantic Anomaly and
intervals of source occultation by the Earth were excluded. The time
entries in SuperAGILE event files are in the on-board time reference
system (Coordinated Universal
Time) and were converted in Terrestrial Dynamical Time before being processed and
analysed by the same procedures used for the GRID
(see \S~\ref{timinganalysis}).
A complete
analysis of the X-ray observations of Crab with SuperAGILE is beyond the
scope of this work: it will be presented in a future paper, as well as
the search for the Crab pulsed signal in the \emph{AGILE}
mini-calorimeter data.

\section{Gamma-ray timing procedures}
\label{timinganalysis}

In this section we will describe the timing procedures we have adopted.
They have been implemented with two aims: to verify the timing performances
of {\em AGILE} and to maximize the quality of the detection of the
four targets.

\emph{AGILE} on-board time is synchronized to Coordinated Universal
Time (UTC) by Global Positioning System (GPS) time sampled at a rate
of 1 Hz. Arrival time entries in \emph{AGILE} event lists files are
then corrected to Terrestrial Dynamical Time (TDT) reference system at
ground segment level. In order to perform timing analysis, they have
also to be converted to Barycentric Dynamical Time (TDB) reference and
corrected for arrival delays at Solar System Barycenter (SSB). This
conversion is based on the precise knowledge of the spacecraft
position in the Solar System frame. To match instrumental microsecond
absolute timing resolution level, the required spacecraft positioning
precision is $\lesssim$0.3 km. This goal is achieved by the
interpolation of GPS position samples extracted from telemetry packets. 
Earth position and velocity with respect to SSB are then
calculated by JPL planetary ephemeris DE405. All the above barycentric 
corrections are handled by a dedicated program (implemented in the 
\emph{AGILE} standard data reduction pipeline) on the event list extracted 
according to the criteria described in \S~\ref{datareduction}.

Pulsed signals in GRID gamma-ray data cannot be simply found by
Fourier analysis of the photon SSB arrival times, since the pulsar
rotation frequencies are 4--5 order of magnitude higher than the
gamma-ray pulsars typical count-rates (10--100 counts/day). The
determination of the pulsar rotational parameters in gamma-ray must
then start from an at least approximate knowledge of the pulsar spin
ephemeris, provided by observations at other wavelengths.

Standard epoch folding is performed over a tri-dimensional grid
centered on the nomimal values of the pulsar spin frequency $\nu_0$
and of its first and second order time derivatives, $\dot{\nu_0}$ and
$\ddot{\nu_0},$ as given by the assumed (radio or X--ray) ephemeris at
their reference epoch $t_0= \rm PEPOCH$. The axes of the grid are explored
with steps equal to 1/T$_{span},$ 2/T$_{span}^2$ and 6/T$_{span}^3$
respectively (all of them oversampled by a factor 20), where
T$_{span}$ is the time span of the gamma-ray data. For any assigned
tern $[\nu;~\dot\nu;~\ddot\nu],$ the pulsar phase $\Phi^{*}$ associated to
each gamma-ray photon is determined by the expression:
\begin{equation}
\label{phasing1}
\Phi^{*}=\Phi_{0}+\nu\Delta t+\frac{1}{2}\dot{\nu}\Delta
t^2+\frac{1}{6}\ddot{\nu}\Delta t^3
\end{equation}
\noindent
where $\Delta t=t-t_0$ is the difference between the SSB arrival time
$t$ of the photon and the reference epoch $t_0$ of the ephemeris and
$\Phi_0$ is a reference phase (held fixed for all the set). A
light-curve is formed by binning the pulsar phases of all the photons
and plotting them in a histogram.  Pearson's $\chi^2$ statistic is
then applied to the light-curves resulting from each set of spin
parameters, yielding the probabilities of sampling a uniform
distribution. These probabilities ${\cal{P}}(\nu;~\dot\nu;~\ddot\nu)$
are then weighted for the number ${\cal{N}}_{\rm st}$ of steps over
the grid which has been necessary to reach the set
$[\nu;~\dot\nu;~\ddot\nu]$ starting from
$[\nu_0;~\dot\nu_0;~\ddot\nu_0].$ The maximum value over the grid of
${\cal{S}}=1- {\cal{N}}_{\rm st}{\cal{P}}(\nu;~\dot\nu;~\ddot\nu)$
finally determines which are the best gamma-ray rotational parameters
for the target source in the surroundings of the given ephemeris. Of
course, the higher the value of ${\cal{S}}$, the higher is the
statistical significance of a pulsating signal. We note that this
approach allows to avoid any arbitrariness in the choice of the range
of the parameters to be explored, which otherwise can affect the
significance of a detection.

Due to the brightness of
the sources discussed in this paper, period folding around the
extrapolated pulsar spin parameters obtained from publicly available
ephemerides (ATNF Pulsar Catalog,\footnote{See http://www.atnf.csiro.au/research/pulsar/psrcat/.}
\citealt{manchester05}, and Jodrell Bank Crab ephemeris
archive,\footnote{See http://www.jb.man.ac.uk/$\sim$pulsar/crab.html.}
\citealt{lyne93}) or from recent literature led us to
firmly detect ($>$$5\sigma$) the pulsations for all the four targets
with a reasonable number of trials ($\lsim$1,000). However, for three
of them (Crab, Vela and PSR B1706$-$44), the best gamma-ray spin
period fell outside the $3\sigma$ uncertainty range of the adopted
radio ephemeris, and the detection significance ${\cal{S}}$ resulted
lower than that derived using contemporary timing solutions
(accounting for the effects of timing noise and/or occasional glitches),
provided by dedicated radio observation campaigns (see \S~\ref{sec:radio}).
As expected, for steady and older pulsars, such as
Geminga, the availability of contemporary rotational parameters
turned out to be less important; even using a few-year-old ephemeris, the
pulsed signal could be detected within only $<$100 period search trials
around the extrapolated X-ray timing solutions.

An additional significant improvement (see \S~\ref{section-tctr}) in
the detection significance has been obtained by accounting also for the
pulsar timing noise in the folding procedure. This exploits a tool
of TEMPO2 \citep{hobbs06,edwards06}, namely the possibility of fitting
timing residuals with a polynomial harmonic function $\Delta R$ in
addition to standard positional, rotational and (when appropriate)
binary parameters \citep{hobbs04}:
\begin{equation}
\label{phasing2}
\Delta R(\Delta t)=\sum_{k=1}^{N}a_k\sin(k\omega\Delta
t)+b_k\cos(k\omega\Delta t)
\end{equation}
\noindent
where $N$ is the number of harmonics (constrained by precision
requirements on radio timing residuals, as well as by the span and the
rate of the radio observations), $a_k$ and $b_k$ are the fit
parameters (i.e. the $\it{WAVE}$ terms in TEMPO2 ephemeris files), and
$\omega=2\pi(T_{radio}(1+4/N))^{-1}$ is the main frequency
(i.e. $\it{WAVE_{OM}}$ in TEMPO2 ephemeris files) related to the radio
data time-span $T_{radio}$.

If the spin behavior of the target is suitably sampled, the harmonic
function $\Delta R$ can absorb the rotational irregularities of the
source, in a range of timescales ranging from $\sim$$T_{radio}$ down to
about the typical interval between radio observations.  As an example, the
peak-to-peak amplitude of the $\Delta R$ fluctuations for Crab related
to the radio monitoring epochs (54,074--54,563 MJD) covering our
\emph{AGILE} observations are of the order of $\sim$1 ms,
corresponding to a phase smearing $>$0.03, a value significantly
affecting the time resolution of a $>$50 bins light-curve. Under the
assumption that the times of arrival of the gamma-ray photons are
affected by timing noise like in the radio band, gamma-rays folding
can properly account for $\Delta R$, extending the relation
\ref{phasing1} to:

\begin{eqnarray}
\label{phasing3}
\Phi & = & \Phi^{*}+ (\nu+\dot{\nu}\Delta
t+\frac{1}{2}\ddot{\nu}\Delta t^2)\Delta R
 \nonumber\\
& \simeq & \Phi_{0}+\nu\Delta t(1+\frac{\Delta R}{\Delta t})+\frac{1}{2}\dot{\nu}\Delta
t^2+\frac{1}{6}\ddot{\nu}\Delta t^3
\end{eqnarray}

As reported in \S~\ref{section-tctr}, this innovative phasing
technique improves significantly the gamma-ray folding accuracy for
young and energetic pulsars, especially when using long data spans,
like those of the \emph{AGILE} observations.  Of course, the
implementation of this procedure requires radio observations
covering the time span of the gamma-ray observations
making the radio monitoring described in \S~\ref{sec:radio}
all the more important.

\section{Timing calibration tests and results}
\label{section-tctr}

\begin{figure} 
\resizebox{\hsize}{!}{\includegraphics[angle=00,width=15cm]{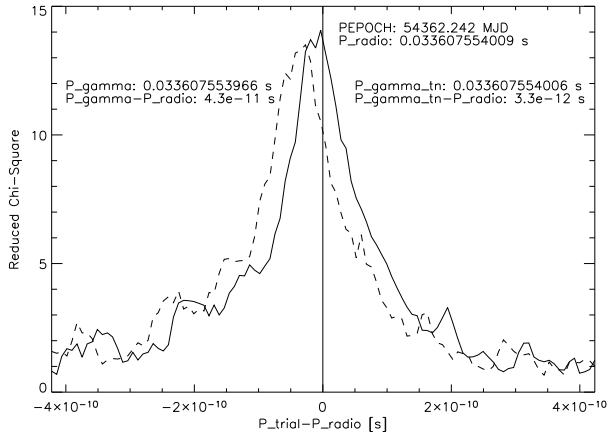}}
\caption{\label{redchi} Period search result for the Crab pulsar (period
trials vs.  $\chi^2$ Pearson statistics). The radio period (vertical
line at $P_{\rm trial}-P_{\rm radio}=0$) is 33.607554009(4) ms
($PEPOCH=54,362.242$ MJD).  Dashed line is obtained from standard folding
period trials, while continuos line is from the folding technique
accounting for timing noise (see \S~\ref{sec:radio}). The new method
allow us to perfectly match the radio period in gamma rays ($P_{\rm
gamma\_tn}-P_{\rm radio}\sim3\times10^{-12}$ s). It is worth noting that
in the considered data span (54,305--54,406 MJD) the period search
resolution ($r_{\rm Crab}\sim2\times10^{-12}$ s) is about an order of
magnitude higher than the 1$\sigma$ error on the radio period.}
\end{figure}

\begin{figure}  
\resizebox{\hsize}{!}{\includegraphics[angle=00,width=15cm]{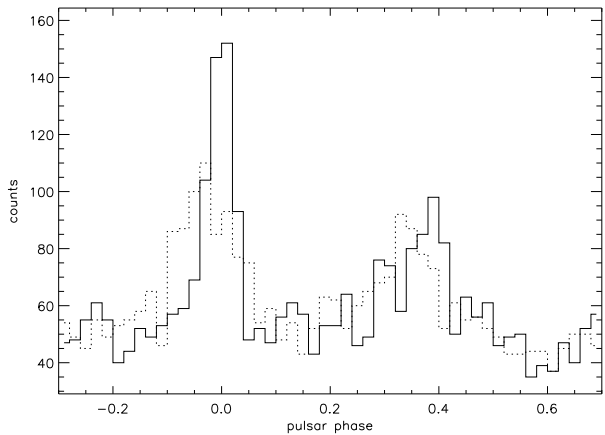}}
\caption{\label{crabcompare} The Crab light-curve (50 bin)
corresponding to the data span 54,305--54,406 MJD
obtained by folding including $\Delta R$ terms compared with that
obtained neglecting timing noise (dashed line).
In observing periods strongly affected by timing noise, the smearing
effects reduce detection
significance and observed pulsed counts (see text).}
\end{figure}

In order to verify the performances of the timing analysis procedure
described in \S~\ref{timinganalysis}, a crucial parameter to check is
the difference between pulsar rotation parameters derived from radio,
X-ray and gamma-ray data.  Figure \ref{redchi} shows the \emph{AGILE}
period search result for the Crab pulsar (corresponding to the MJD
54,305--54,406 observations, significantly affected by timing noise,
as shown in Figure \ref{fig:residual}). The implementation of the folding
method described in section \S~\ref{timinganalysis} (including timing
noise corrections as given by Equation [\ref{phasing3}]) allowed for a
perfect match between the best period resulting from gamma-ray data
and the period predicted by the radio ephemeris with
discrepancies $\Delta P_{\rm Crab}\sim3\times10^{-12}$ s, comparable
to the period search resolution $r_{\rm Crab}\sim2\times10^{-12}$
s. This represents also an ultimate test for the accuracy of the
on-board \emph{AGILE} Processing and Data Handling Unit
\citep[PDHU;][]{argan04} time management (clock stability in
particular) and on-ground barycentric time correction
procedure. Standard folding without $\Delta R$ term implies
radio-gamma period discrepancies one order of magnitude higher
($\Delta P_{\rm Crab}\sim4\times10^{-11}$ s). Moreover, it lowers also
the statistical significance of the detection and effective time
resolution of the light-curve.
For the Crab, the value of the Pearson's $\chi^2$ statistic
introduced in \S~\ref{timinganalysis} 
(we quote reduced $\chi^2$ values)
goes from $\sim$6.3 (when
using the $\Delta R$ term) down to $\sim$4.2 (ignoring the
$\Delta R$ term) when folding the data into a 50 bins light-curve (see
Figure~\ref{crabcompare}). Obviously,
ignoring timing noise in the folding process would yield discrepancies
(and light curve smearing) which are expected to grow when considering
longer observing time span. Thus the contribution of timing noise
should be considered both in high-resolution
timing analysis and in searching for new gamma-ray pulsars. The same
analysis applied to Vela and PSR~B\,1706--44 -- much less affected
than Crab by timing noise in the considered data span -- led to
similar results for the period discrepancies ($\Delta P_{\rm
Vela}\sim8\times10^{-12}$ s, $\Delta P_{\rm
B1706}\sim3\times10^{-11}$, to be compared with the period search
resolution $r_{\rm Vela}\sim9\times10^{-12}$ s and $r_{\rm
B1706}\sim2\times10^{-11}$ s).


For the radio-quiet Geminga pulsar we used X-ray ephemeris obtained from
\emph{XMM-Newton} data (see \S~\ref{sec:radio}) as a starting
point for the period search. Due to the stability of the spin
parameters of this relatively old pulsar, not significantly affected
by timing noise, $\it{WAVE}$ parameters are not required for the
folding process. The X-ray versus gamma-ray period discrepancy was
$\Delta P_{\rm Gem}\sim9\times10^{-12}$ s, whereas
$r_{\rm Gem}\sim7\times10^{-11}$ s.  We here
note that the frequency resolution 1/T$_{\rm span}\sim10^{-7}$ Hz of
\emph{AGILE} is about one order of magnitude better than that
corresponding to a single \emph{XMM-Newton} exposure (lasting $\lsim100$
ks), but the few year long X-ray data span implies an overall much
better effective resolution 1/T$_{\rm XMM}\sim5\times10^{-9}$ Hz for the
\emph{XMM-Newton} data.

\begin{figure} 
\resizebox{\hsize}{!}{\includegraphics[angle=00,width=15cm]{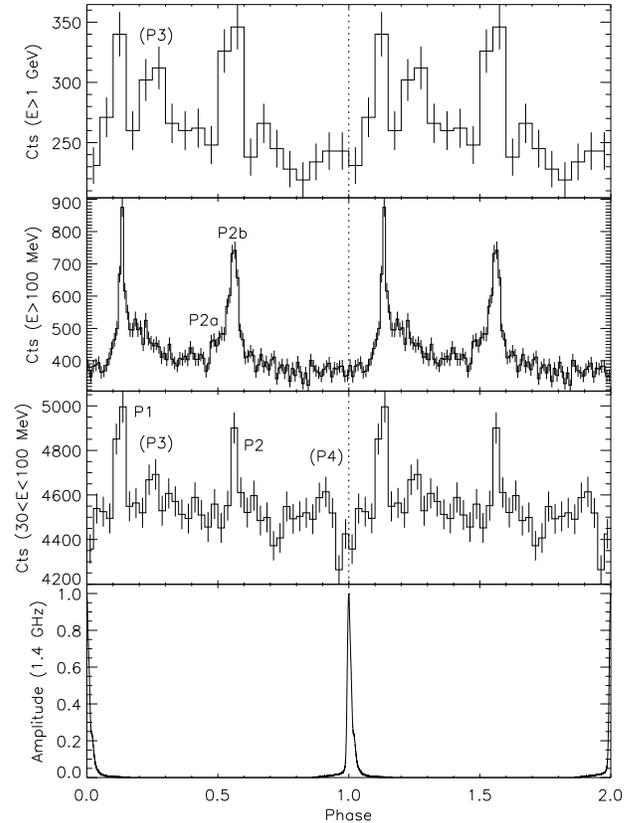}}
\caption{\label{vela} Vela pulsar light-curves ($P\sim89.3$ ms) for
different energy bands ($E<100$ MeV, 40 bins, resolution: $\sim$2.2 ms;
$E>100$ MeV, 100 bins, resolution: $\sim$0.9 ms; $E>1$ GeV, 20 bins,
resolution: $\sim$4.5 ms; G+L class events) obtained by integrating all available
post-glitch data (54,320--54,561 MJD).  The radio ephemeris and
the 8,192 bins light-curve
(bottom panel) are obtained by the analysis of
$\sim$4,100 ToAs observed at Mt.~Pleasant Radio Observatory in
Tasmania (radio observation interval 54,157--54,548 MJD).}
\end{figure}

\begin{figure} 
\resizebox{\hsize}{!}{\includegraphics[angle=00,width=15cm]{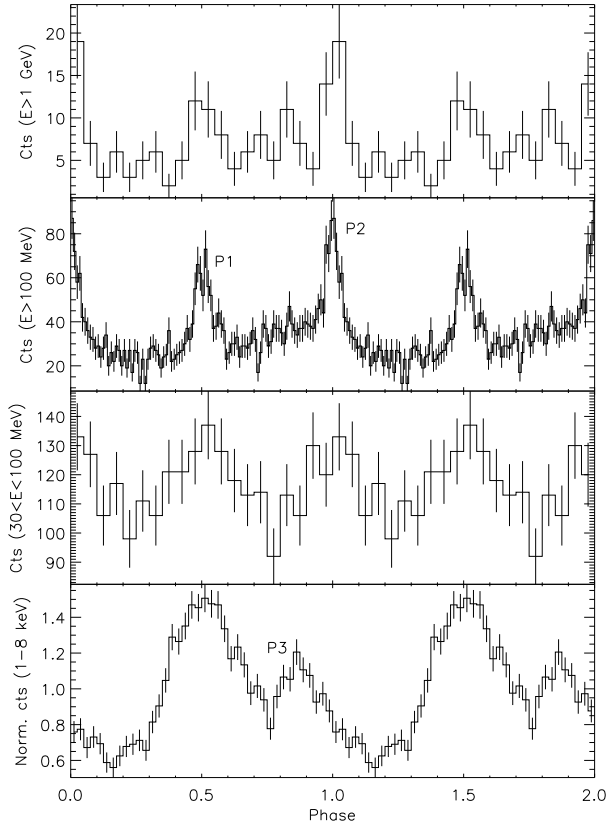}}
\caption{\label{geminga} Geminga light-curves ($P\sim237.1$ ms) for
different energy bands ($E<100$ MeV, 20 bins, resolution: $\sim$11.8 ms;
$E>100$ MeV, 100 bins, resolution: $\sim$2.4 ms; $E>1$ GeV, 20 bins,
resolution: $\sim$11.8 ms; G class events).
The X-ray ephemeris and the 1--8 keV 40 bins light-curve
(bottom panel) are obtained by the analysis of \emph{XMM-Newton} 
data (observation interval 52,369--54,534 MJD).}
\end{figure}

\begin{figure} 
\resizebox{\hsize}{!}{\includegraphics[angle=00,width=13cm]{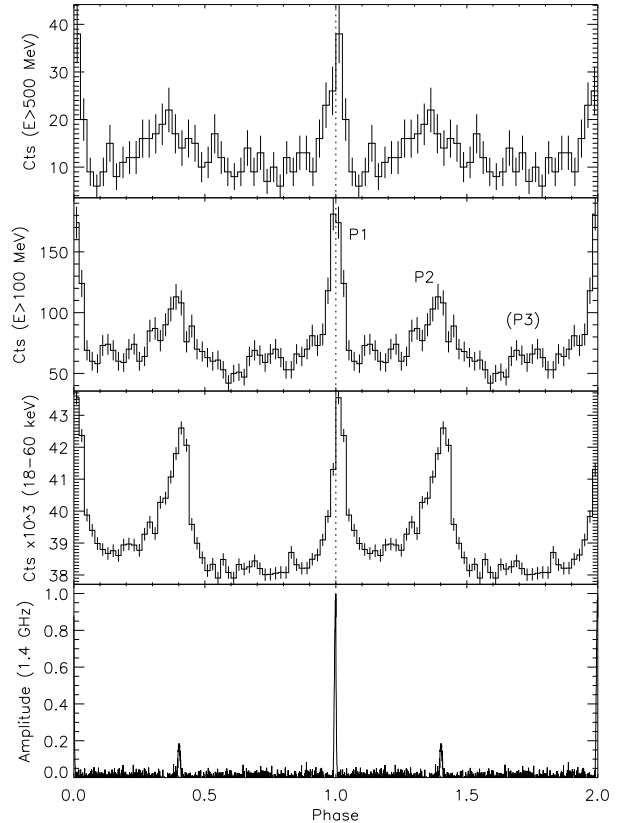}}
\caption{\label{crab} Crab pulsar light-curves ($P\sim33.1$ ms) for
different energy bands ($E>100$ MeV, 50 bins, resolution: $\sim$0.7 ms;
$E>500$ MeV, 40 bins, resolution: $\sim$0.8 ms; G class events) obtained by integrating all available
 data presented in Table \ref{counts}.  The X-ray (18--60 keV)
SuperAGILE 50 bins light-curve is obtained from a $\sim$41 ks observation
taken at $\sim$54,361 MJD. The radio ephemeris and
the 2048 bins light-curve
(bottom panel) are obtained by the analysis of
334 ToAs observed by Jodrell Bank and Nan\c cay radio telescopes (observation
interval 54,078--54,566 MJD).}
\end{figure}

\begin{figure} 
\resizebox{\hsize}{!}{\includegraphics[angle=00,width=15cm]{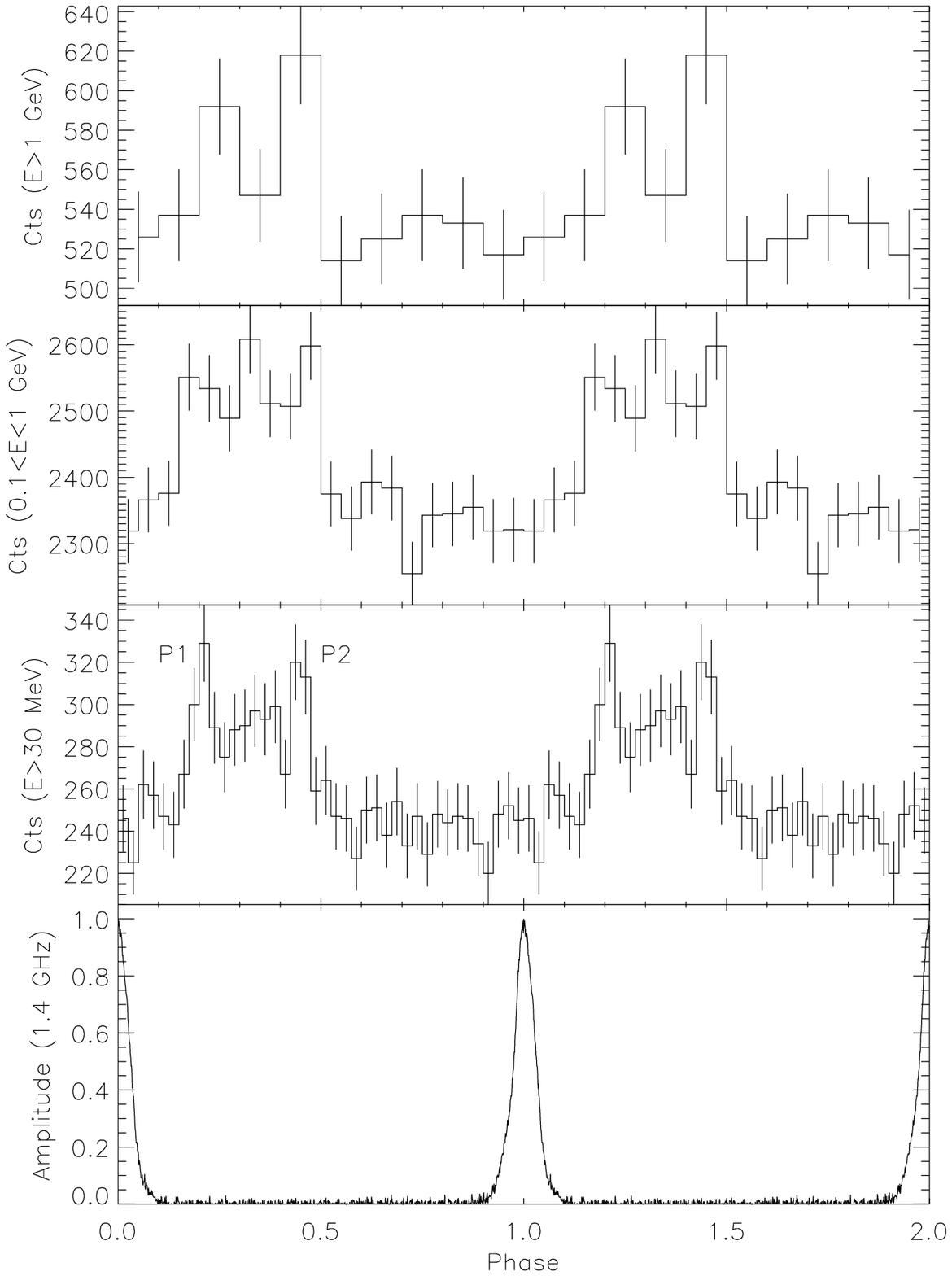}}
\caption{\label{J1709} PSR B\,1706--44 light-curves ($P\sim102.5$ ms)
for different energy bands ($E>30$ MeV, G class events, 40 bins, resolution: $\sim$2.6 ms;
$0.1<E<1$ GeV, G+L class events, 20 bins, resolution: $\sim$5.1 ms; $E>1$ GeV, G+L class events, 10 bins,
resolution: $\sim$10.2 ms) obtained by integrating all available
 data presented in Table \ref{counts}. The radio ephemeris and
the 1024 bins light-curve
(bottom panel) are obtained by the analysis of
20 ToAs observed by Parkes radio telescope (observation
interval 54,220--54,562 MJD).}
\end{figure}

The resulting gamma-ray light-curves, covering different energy ranges
for the four pulsars, are shown in Figures \ref{vela}, \ref{geminga},
\ref{crab}, and \ref{J1709}.  The pulsed flux was computed considering
all the counts above the minimum of the light-curve, using
the expression $PF=C_{\rm{tot}}-n N_{\rm{min}}$ and its associated
error $\sigma_{PF}=(C_{\rm{tot}}+n^2 \sigma^2_{N_{\rm{min}}})^{1/2}
\simeq n (N_{\rm{min}})^{1/2}$, where $C_{\rm{tot}}$ are the total
counts, $n$ is the number of bins in the light-curve and
$N_{\rm{min}}$ are the counts of bin corresponding to the minimum.
This method is ``bin dependent'', but reasonable different choices of
both the number of bins (i.e. $n>10$) and of the location of the bin
center (10 trial values were explored for each choice of $n$) do not
significantly affect the results.  Several models (polar cap, slot
gap) predict that gamma-ray pulsar emission is present at all phases.
When enough counts statistics will be available,
(e.g. $\sim$10,000 counts for the Crab pulsar),
it will be possible to estimate the unpulsed emission (due to
the pulsar plus a possible contribution from the pulsar wind nebula)
by considering the difference among the total source flux obtained by
a likelihood analysis on the images and the pulsed flux estimated with the above
method.

Pulsed counts and related Pearson statistics for the four pulsars are
reported in Table \ref{counts2} (for the standard event extraction
parameters as in Table \ref{counts}).  The resulting fluxes (Pulsed
Counts/Exposure) are consistent with those reported in the EGRET
Catalogue \citep{hartman99}.  We note that source-specific extraction
parameters (event class, source position in the field of view, energy
band etc.) which maximize reduced $\chi^2$ can significantly improve
detection significance. For example, including only event class G in
the timing analysis of Crab and Geminga halves the number of pulsed
counts while doubling the $\chi^2$ values.

Despite the very satisfactory matching of the pulsar spin
parameters found in radio (or X-ray) and gamma-ray (supporting the
clock stability and the correctness of the SSB transformations),
possible systematic time shifts in the \emph{AGILE} event lists could
be in principle affecting phasing and must be checked.  For example, an
hypothetical constant discrepancy of t$_{err}$ of the on-board time
with respect to UTC would result in a phase shift
$\Phi_{err}=(t_{err}~\rm{mod}~P)/P$, where $P$ is the pulsar period.
The availability of radio observations bracketing the time span of the
gamma-ray observations (or of X-ray observations very close to the
gamma-ray observations for the case of Geminga) allowed us to also
perform accurate phasing of multi-wavelength light-curves. In doing
that, radio ephemeris reference epochs were set to the main peak of
radio light-curves at phase $\Phi_{peak}=0$.  In view of
Equations (\ref{phasing2}) and (\ref{phasing3}), this is achieved by
setting $\Phi_0=-\nu\sum_{k=1}^{N}b_k$ (typically $\Phi_0<10^{-2}$).
We found that the phasing of the \emph{AGILE} light-curves of the four
pulsars (radio/X-rays/gamma-ray peaks phase separations) is consistent
with EGRET measurements 
\citep[][see \S~\ref{discuss} for details]{fierro98,tbb96,jackson05} 
implying no evidence of systematic errors in absolute timing with an 
upper limit $t_{err}<1$ ms.

Comparison with the SuperAGILE light-curve peak (see \S~\ref{sec:radio})
is also interesting (SuperAGILE on-board time processing is more complex
than that of the GRID). The Crab SuperAGILE light-curve (Figure \ref{crab}) 
was produced with the same folding method reported in \S~\ref{timinganalysis}, 
yielding $63,700\pm8,700$ pulsed counts ($\sim$3\% of the total counts 
including background). Inspection of Figure \ref{crab} shows that the 
X-ray peaks are aligned with the $E>100$ MeV data within 
$\Delta\phi\sim400$ $\mu$s (a value obtained fitting the peaks with Gaussians) 
providing an additional test of the \emph{AGILE} phasing accuracy.

The effective time resolution of \emph{AGILE} light-curves results
from the combination of the different steps involved in the processing
of gamma-ray photon arrival times. The on-board time tagging accuracy
is a mere $\sim$1 $\mu$s, with negligible dead time. For comparison, 
the corresponding EGRET time tagging accuracy was $\sim$100 $\mu$s. The 
precise GPS space-time positioning of \emph{AGILE} spacecraft 
\citep{argan04} allows for the transformation from UTC to Solar System 
barycenter time-frame (TDB) with only a moderate loss ($\lesssim$10 $\mu$s) 
of the intrinsic instrumental time accuracy. The innovative folding
technique described in \S~\ref{timinganalysis}, accounting for pulsar
timing noise, also reduces smearing effects in the light-curves, fully
exploiting all the information from contemporary radio observations.
In summary, the effective time resolution of the current \emph{AGILE}
pulsar light-curves (and then multi-wavelength phasing accuracy
assuming $t_{err}=1$ $\mu$s) is mainly limited by the available count
statistics and can be estimated by:
\begin{equation}
\Delta t={{P}\over{N}}={{\sigma^2(C_p+2B)}\over{C_p^2}}
\end{equation}
\noindent
where $P$ is pulsar period, $N$ is the number of bins in the
light-curve histogram, $\sigma$ is the signal-to-noise ratio, $C_p$
are the pulsed counts and $B$ are the background counts.  In order to
keep the average signal-to-noise ratio of light-curve bins (during the
on-pulse phase) at a reasonable level ($>$3\,$\sigma$), the resulting
effective time resolution is constrained to 200--500 $\mu$s.  At
present, the best effective time resolution ($\sim$200 $\mu s$) is
obtained for the 400-bin light-curve of Vela (G+L class selection)
although a 100-bins light-curve (Figure \ref{vela}) is better suited
to study pulse shapes and to search for possible weak features. The 
effective time resolution will obviously improve with exposure time 
$\Delta t \propto T_{exp}^{-1}$ and a resolution $\lsim$50 $\mu$s is 
expected after two years of \emph{AGILE} observations of Vela.

\section{Discussion}
\label{discuss}

With about 10 years since the last gamma-ray observations
of Crab, Vela, Geminga \citep{fierro98} and B\,1706--44
\citep[][]{tbb96} by CGRO, the improved time resolution of
\emph{AGILE} and the much longer observation campaigns in progress are
now offering the possibility of both to search for new features in the
shape of the light-curves of these gamma-ray pulsars and to
investigate the possible occurrence of variations in the gamma-ray
pulsed flux parameters.

After nine months of observations in the frame of the Science
Verification Phase (2007 July--November) and of the Scientific
Pointing Program AO\,1 (pointings 1--7, 2007 December 1--2008 April 10), 
\emph{AGILE} reached an exposure (E$>$100 MeV) of the Vela region $\gtrsim$$10^9$
cm$^2$ s ($\sim$10,000 pulsed counts from Vela), comparable to that of
the nine-year life of EGRET (although \emph{AGILE} data have a higher
residual particle background), and an even better exposure ($1.5\times10^9$ cm$^2$ s) in
the core region of the Galactic Plane ($l=310\degr$--$340\degr$)
corresponding to the Southern Hemisphere . In fact, \emph{AGILE}
observed $\sim$2,400 pulsed counts from PSR B1706$-$44 up to date, a
factor 1.5 better counts statistics than EGRET for this source.
For Crab and Geminga, an exposure level comparable with that obtained
by EGRET will be reached at the end of the AO\,1 pointing number 15 in the
Anti-Center region (October, 2008).

\subsection{The gamma-ray light-curves}
\label{lc}

The plots shown in Figures \ref{vela}, \ref{geminga}, \ref{crab}, and 
\ref{J1709} allow us to start assessing new features in 
gamma-ray pulsar light-curves. Narrower and better resolved main peaks 
are revealed, together with previously unknown secondary features, to be 
confirmed when more count statistics (and an improved particle background 
filtering) will be available.

The Vela light-curves for different energy bands are shown in Figure
\ref{vela}.  A Gaussian fit to the Vela main peak (P1) at $E>100$ MeV
provides a FWHM of $0.018\pm 0.002$ centered at $\phi=0.1339\pm0.0007$
in phase consistent with the EGRET observations (Kanbach et al., 1994).
The apex of the main peak is resolved by \emph{AGILE} with a width of
$\sim$0.8 ms and its apparent trail ($\phi=0.13$--0.3) could be
due to the occurrence of one or more secondary peaks. In
fact, marginal evidence of a relatively narrow lower peak (P3) at
$\phi\sim0.25$ is present in the $E<$100 MeV (3.3$\sigma$ fluctuation
with respect to the average interpulse rate) and in the $E>$1 GeV (3.8$\sigma$)
light-curves. P3 is located at the phase of the optical peak 1, and also at
the phase of a bump predicted in a two-pole caustic model
\citep{dyks04}, due to overlapping field lines from opposite poles
near the light cylinder. In the outer gap model, this bump is the
first peak in the light-curve, and also comes from very near the light
cylinder. The peak at $\phi=0.5$--0.6 (P2) in the $E>100$
MeV light-curve cannot be satisfactorily fit with a single Gaussian or
a Lorentzian curve 
($\chi^2_r>3$), due the possible presence of a bump at $\phi\sim0.5$
(P2a). A fit with two Gaussian provides $\phi=0.560\pm0.001$ (FWHM
$0.031 \pm 0.003$) and $\phi=0.49\pm0.01$ for the phases of the major
peak (P2b) and lower peak apex (P2a), respectively. The phase
separation between the main gamma-ray peaks $\Delta
\phi=0.426\pm0.002,$ as well as that between the gamma-ray and the
radio peak, are unchanged since EGRET observations \citep{kanbach94,ramanamurthy95}. 
In the $E<100$ MeV band a secondary peak+valley structure (P4) appears at
$\phi\sim0.9$ ($\sim$4$\sigma$ level).  It is worth noting its
symmetric position around the radio peak with respect to main peak
(P1) and a possible correlation of P4 with features seen in the X-ray
light-curves \citep{manzali07}; P4 also coincides with peak number 3 of the
\emph{RossiXTE} light curve \citep{hsg02}.

Geminga shows an $E>100$ MeV light-curve (Figure \ref{geminga}) with
properties similar to those of Vela. Apart from the major peaks P2
($\phi=0.999\pm0.002$, FWHM $0.062\pm0.008$) and P1 (with apex
$\phi=0.507\pm0.004$, FWHM $0.08 \pm 0.01$), secondary peaks are seen
trailing P1 and leading P2: the main peak's P2 leading trail
($\phi=0.8$--1) could be possibly associated to unresolved multiple
peaks while P1 displays a ``bump'' at
$\phi\sim0.55$--0.6. P1 seems to be characterized by a double structure:
a fit with a simple Gaussian yields $\chi^2_r\sim1.2$ to be
compared with a double Gaussian model providing $\chi^2_r<1$.  This
feature seems still present even considering different observation
block separately.  The main-peak separation ($\Delta
\phi=0.508\pm0.007$ at $E>100$ MeV) is greater than the value obtained for Vela and it
decreases slightly with energy. The 2--10 keV X-ray light-curve shows
a peak (P3) in correspondence with a possible excess in the hard
gamma-ray band and a broad top-hat shaped feature partially
overlapping in phase with P1.

The Crab light-curve for $E>100$ MeV (Figure \ref{crab}) shows a
previously unknown broad feature at $\phi=0.65-0.8$ (P3), in addition
to the main peaks P1 and P2 ($\phi_{P1}=0.999\pm0.002,$ FWHM $0.054
\pm 0.005$; $\phi_{P2}=0.382\pm0.008,$ FWHM $0.14 \pm 0.04$). The
probability of P3 being a background fluctuation is of $\sim$$10^{-4}$
(3.7$\sigma$). P2 could be possibly resolved in two sub-peaks in
future longer observations.  P3 is coincident with the feature HFC2
that appears in the radio profile above 4 GHz \citep{moffett96}.  From
the polarization of this component, \citet{moffett99} suggest that
this peak may come from a lower emission altitude, near the polar
cap. P3 is actually at a phase that could plausibly come from low
altitude cascades in the slot gap model
\citep[][Figure~2]{muslimov04}, the pairs from which may be also
causing the HFC2 radio component(s), while P1 and P2 comes from the
high-altitude slot gap.

According to the observations of \emph{SAS-2}, \emph{COS B} and
\emph{CGRO}/EGRET, the ratio P2/P1 of the main peak intensities could
present a variability pattern (possibly ascribed to
the nutation of the neutron star) that can be fit with a sinusoid with a
period of $\sim$13.5 years \citep[][]{kanbach90,ramanamurthy95}
although this is not required by EGRET data alone \citep{tompkins97}.  We
observed a P2/P1 intensity ratio $0.66\pm0.10$ in good agreement with
the value of $\sim$0.59 predicted for 54,350 MJD (for the energy range
50 MeV--3 GeV).
Unfortunately, our P2/P1 value is
similar - within the errors - to the EGRET determination ($\sim$0.5):
then an unambiguous assessment of the origin of this possible phenomenology
will require measurements close to the epoch (56,150 MJD)
corresponding to the predicted maximum or the intensity ratio P2/P1
($\sim$1.4).  Variability should also be invoked to explain the
possible detection of P3, which was never seen before in the EGRET
database in spite of an overall exposure comparable to that reached by
\emph{AGILE} so far.  We note that the main peak intensity ratios
computed for Vela ($\rm{P2/P1}= 0.91\pm0.07$) and Geminga ($\rm{P2/P1}=
0.8\pm0.1$) do not yield evidence of significant variations with
respect to past observations \citep{ramanamurthy95}.

The \emph{AGILE} light-curves of PSR~B\,1706$-$44 are shown in Figure
\ref{J1709}. The broad-band light-curve ($E>30$ MeV) clearly shows two
peaks ($\phi_{P1}=0.211\pm0.007,$ $\phi_{P2}=0.448\pm0.005$) bracketing
considerable bridge emission (contributing to $>$50\% of the pulsed
counts) while in the 0.1--1 GeV range the peaks
cannot be discerned from the bridge emission and the pulsar profile presents
an unresolved broad
($\Delta \phi=0.3-0.4$) single peak.
PSR B\,1706--44
is a young ($\sim2\times10^4$ yr) and energetic ($3.4\times10^{36}$
erg s$^{-1}$) 102.5 ms pulsar \citep{jlm92} with emission properties similar
to Vela \citep{becker02}.  Double peaked PSR~B\,1706--44 is then in
fact ``Vela-like'' not only energetically, but also with respect to
the offset between the maxima of the high energy and the radio
profiles, with neither of the two gamma-ray narrow pulses aligned to
the radio peak.

\emph{AGILE} allowed for a long monitoring of gamma-ray pulsar
light-curves shapes. 
We carefully looked for possible lightcurve variations by KS tests
(two-dimensional Kolmogorof-Smirnoff test [KS]; \citealt{peacock83,fasano87}).
For each pulsar, different gamma-ray light-curves (with 10, 20
and 40 bins) were obtained grouping contiguous data set and requiring
at least 30 counts bin$^{-1}$ (300--1000 counts for each light-curve). Each
light-curve was compared by KS test with each other and with the
average shape corresponding to the entire data set. No pulse shape
variation was detected with a significance $>$3\,$\sigma$ on
timescales ranging from 1 days (Vela) to few months (Crab,
Geminga and PSR~B\,1706$-$44). 

\subsection{Implications for the emission models}
\label{models}

Pulsars derive their emitting power from rotational energy loss owing
to the relativistic acceleration (up to Lorentz factor
$\Gamma\sim10^5$--$10^7$) of charged particles by very high electric
potentials induced by the rotating magnetic fields. The charge density
that builds up in a neutron star magnetosphere is able to short out
the electric field parallel to the magnetic field everywhere except a
few locations of non-force-free ``gaps''. It is unclear whether these
acceleration regions can form in the strong field near ($\lesssim$1
stellar radii) the neutron star surface (polar cap/low-altitude slot
gap model: \citet{daugherty96} and \citet{muslimov03}; high-altitude
slot gap model: \citet{muslimov04} and \citet{harding08}), in the
outer magnetosphere near the speed of light cylinder (outer gap model:
\citet{cheng86,romani96,hirotani01,takata07}) or even beyond in the
wind zone \citep{petri05}.

In polar cap/low altitude slot gap models, gamma-rays result from
magnetic pair cascades
induced by curvature or
inverse-Compton photons. The spectrum is dominated by synchrotron
radiation of pairs at lower energies ($\lesssim$1 GeV) and by
curvature radiation at higher energies ($>$1 GeV). In the
high-altitude slot gap, gamma-rays result from curvature radiation and
synchrotron radiation of primary electrons (no cascade). In outer gap
models, gamma-rays result from photon-photon pair cascades induced by
curvature radiation. The spectrum is dominated by pair synchrotron
radiation below 20 MeV, curvature radiation above 20 MeV and inverse
Compton at 1--10 TeV.

The size and spectrum of the emitting regions are then directly
related to the intensity and location in the magnetosphere of the
accelerating electric fields. Strong fields imply thin accelerator
gaps, while weaker fields are associated to thick gaps: the
acceleration zone grows bigger as the particle must accelerate over
larger distance to radiate pair-production photons. On an
observational point of view, it is in turn expected that acceleration
gap sizes are related to the width of light-curve peaks. The highly
relativistic particles emit photons at very small ($\sim$$1/\Gamma$)
angles to the open magnetic field lines. The theoretical width of a
light-curve peak associated to an infinitely small gap would be then
$\Delta t = P/(2 \pi \Gamma)$, a value typically smaller than 1
$\mu$s. Therefore, the width of the apex of the peak can be related to
the core gap size. For example, the $\Delta t\sim800$ $\mu$s width of
the Vela pulsar peak (P1) resolved by \emph{AGILE} implies a projected
gap core width of $\sim$1 km for gap height of $\sim$1
stellar radii. Broader peaks as P2 and P3 in the Crab light-curve
involves instead magnetospheric region tens of kilometers long. The
relation of the width of the peaks to the acceleration gap size works
for the low altitude emission models but is complicated by
relativistic effects (aberration, retardation) in the high-altitude
emission models, where the peaks are formed by caustics
\citep{dyks03}. The peak widths will still depend on the width of the
acceleration (outer or slot) gap, but one must perform a comparison
with detailed models to constrain the physical size of the gap.

The likely presence of multiple contiguous peaks as P2a-b in the Vela
light-curve, P1 in Geminga and possibly within the wide P2 broad peak
in Crab could be related to short-term oscillation of gap
locations. Light-curve variations on timescales $\gtrsim$1 days are
anyway excluded by the KS tests described above. Alternatively, the
apparent superimposition of different gaps placed at different heights
in the magnetosphere could be another plausible explanation for
multiple-contiguous peaks, while the hypothesis of jumps in the
acceleration electric fields strength would lack of evident physical
justification. In general, the presence of multiple (contiguous or
not) structures in the light-curves is difficult to fit in a scenario
alternatively involving polar cap, outer gap or wind zone models
exclusively. The particle acceleration may well be simultaneously
occurring in all the regions predicted by these models.
Indeed, a three peak light-curve can still be explained invoking
polar cap gaps alone provided that both polar caps can cross the line-of-sight.

Wherever in the magnetosphere a gap can form, a multi-frequency
emission may occur along an hollow cone (due to magnetosphere
symmetries) or any other suitable surface: light-curve peaks are
generated when the viewing angle from any given location on this
surface to the observer also crosses the gap. The pulse spectrum slope
depends on the accelerating field as well as on the magnetic fields,
which strongly affect the synchrotron emission efficiency and the pair
attenuation along the cascade path: the stronger the magnetic field
is, the steeper the spectrum is. An outer gap can in principle
generate pulse spectra extending up to tens of GeV \citep{zc00}, while
gaps close to surface high magnetic fields can produce steep spectra
\citep[photon index $\lesssim$3;][]{ramanamurthy96} and cut-off in the
tens of MeV range \citep[PSR\,1509--58,][]{harding97}. The absence of
corresponding gamma-ray pulse in phase with radio main peaks in Vela
and PSR~B\,1706--44 could be then ascribed to gap-dependent spectral
slope and not only to beam angle and viewing geometry differences.

The multiplicity and variety of features seen in \emph{AGILE}
light-curves can pave the way to a parameterized standard
model (e.g. with adjustable accelerating electric fields strength,
and location in the magnetosphere) for pulsars gaps and their
corresponding observed high-energy
pulses.  In this perspective, the \emph{AGILE} light-curves time
resolution, currently limited only by the (continuously increasing) source
counts statistics, will eventually yield a pulsar gaps
map by coupling timing analysis and phase resolved spectral analysis.

\subsection{The Vela glitch of August 2007}
\label{velaglitch}

During early \emph{AGILE} observations, Vela experienced a weak
glitch clearly detected in radio as a discontinuity in the pulsar's
spin parameters.

Glitches are small ($\Delta\nu/\nu\sim10^{-9}$--$10^{-6}$) and sudden
($\lesssim$1 day) discontinuous increases in the pulsar frequency,
often followed by a recovery (1--100 days) to the pre-glitch
frequency. About $\sim$6\% of pulsars are known to have shown
glitches\footnote{See e.g. http://www.atnf.csiro.au/research/pulsar/psrcat/.}),
with a higher incidence of events in younger pulsars. This phenomenon
is potentially a very promising tool for probing the physics of the
neutron star interiors \citep{lyne00}.  Although no general consensus
has been reached to date about the origin of the glitches, many models
are based on the exchange of angular momentum between the superfluid
neutron star core and its normal, solid crust \citep{ruderman76,ruderman91,alpar84,aaps84}. 
This angular momentum transfer may excite starquake waves,
propagating toward the neutron star surface.  Since the magnetic field
frozen in the crust is ``shaken'', the resulting oscillating
electromagnetic potential could generate strong electric fields
parallel to the magnetic field, which in turn would accelerate
particles to relativistic energies, possibly emitting a burst of 
high-energy radiation.

Since the first observation of a pulsar glitch in 1969
\citep{radhakrishnan69}, Vela has shown $\sim$10 major glitches.
Due to its large field of view, the quest for possible gamma-ray
bursting behavior due to a glitch is then an effective opportunity for
\emph{AGILE}. Despite the fact that the August 2007 glitch is a weak one
($\Delta\nu/\nu\sim10^{-9}$), it is worthwhile to 
search for a signal in the \emph{AGILE} data.

The characteristic energy of a pulsar glitch can be roughly estimated from the
associated pulsar frequency jump $E_{\rm glitch}=\Delta E_{\rm rot}=4\pi^2
I\nu\Delta\nu$, where $I$ ($\sim$10$^{45}$ g cm$^2$) is the neutron
star momentum of inertia. The corresponding expected gamma-ray counts
would be:
\begin{equation}
C_{\gamma}^{\rm glitch}=\eta{{ E_{\rm glitch}A_{\rm eff}}\over{4\pi d^2
E_{\gamma}}} \simeq 10^{11} \eta {\Delta\nu\over\nu}
\label{eqn:cglitch}
\end{equation}
\noindent
where $\eta=[0,1]$ is the unknown conversion efficiency of the glitch
energy to gamma-ray emission, $d$ ($\sim$0.3 kpc) is the pulsar
distance, $E_{\gamma}$ ($\sim$300 MeV) is the average gamma-ray photon
energy assuming a spectral photon index $\Gamma=-2$, and A$_{\rm eff}$ is
the \emph{AGILE} effective area.

\begin{figure} 
\resizebox{\hsize}{!}{\includegraphics[angle=00,width=15cm]{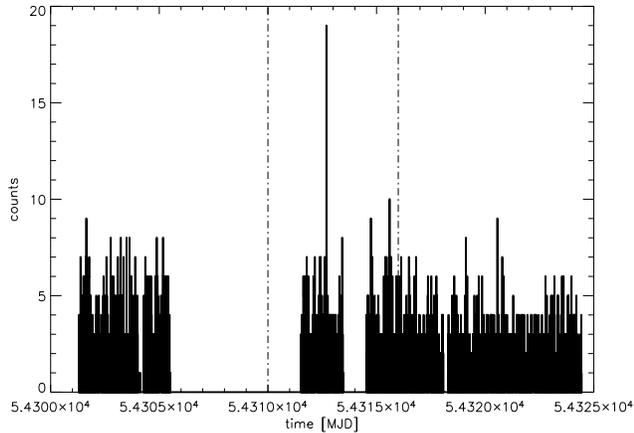}}
\caption{\label{glitch} Unfolded Vela pulsar light-curve (4.5 min binsize, $E>50$ MeV, G-class events).
Dashed lines bracket glitch epoch uncertainty range ($54,313\pm3$ MJD).
A $>$5$\sigma$ count excess at $\sim$54,312.693 MJD could be associated to
gamma-ray bursting emission from the glitch.
}
\end{figure}

Even in the virtual limit assumption that the entire glitch energy could
be driven into gamma-ray emission, a weak glitch with a frequency
shift of ${\Delta\nu/\nu}= 1.3\times10^{-9},$ as that observed in
August 2007, cannot produce a strong signal
($C_{\gamma}^{\rm glitch}<100$--200 counts), if the core fluence is spread
in $\sim$1 day. In fact, no excess on daily timescales was detected,
although for much shorter timescales of 3--6 minutes a
$>$5\,$\sigma$ excess ($\sim$15 counts) in the photon counts is
seen at $\sim$54,312.693 MJD (Figure \ref{glitch}).

On the other hand, stronger Vela glitches, as that of 1988 Christmas
\citep{mcculloch90} with a frequency shift of ${\Delta\nu/\nu} =
2\times10^{-6}$, could in principle produce more significant transient
gamma-ray emission. Typical count-rate from Vela is $\sim$100--200
counts day$^{-1}$, then a fluence of $\gtrsim$1000 counts in $\sim$1-day 
or less from the hypothetical gamma-ray glitch burst should be easily
detectable. According to Equation (\ref{eqn:cglitch}), such a flux could
arise from a glitch with ${\Delta\nu/\nu}\sim 10^{-7}$ (typical Vela
glitch size), converting a relatively small fraction ($\eta\sim0.1$)
of its energy in gamma-rays. The chance occurrence of a strong Vela
glitch in the \emph{AGILE} field of view over three years of mission
operations is of $\sim$20\%.

\section{Conclusions}
\label{conclusions}

\emph{AGILE} collected $\sim$15,000 pulsed counts from known gamma-ray pulsars
during its first 9 months of operations. The \emph{AGILE} PDHU clock
stability, coupled with the exploitation of pulsar timing noise
information,
allows for pulsar period fitting
with discrepancies with respect to radio measurements at the level of
the period search resolution ($\sim$$10^{-12}$ s) over the long
gamma-ray data span ($>$6 months). Thanks to \emph{AGILE} GPS-based high
time tagging accuracy ($\sim$1 $\mu$s), the effective time resolution
of \emph{AGILE} light-curves is limited only by the count statistics
at current level of exposure ($\sim$$1.2\times10^9$ cm$^2$ s for the Vela
region).  The best effective time resolution obtained for Vela
observations is of $\sim$200 $\mu$s for a signal-to-noise
$>$3\,$\sigma$ in the on-pulse light-curve bins. An improved effective
time resolution $\lesssim$ 50$\mu$s is expected after three years of
\emph{AGILE} observations.

\emph{AGILE} multi-wavelength phasing of the four gamma-ray pulsars is
consistent with the results obtained by \emph{EGRET}, although
the high resolution \emph{AGILE} light-curves show narrower and
structured peaks and new interesting features. In particular, a third
peak is possibly detected at $\sim$3.7$\sigma$ level in the Crab light-curve
and several interesting features seem present in the Vela light-curves.

In any case, the highly structured light-curves hint at a complex
scenario for the sites of particle acceleration in the pulsar
magnetospheres, implying different electric gaps with physical
properties probably mostly related to their height above the neutron
star surface. Alternatively, slight spatial oscillations of the gap
locations on timescales $\lsim$1 day could be invoked to explain the
multiple contiguous peaks seen in the light-curves.

The foreseen balance of count statistics at the end of AO\,1 observing
program will allow for phase-resolved spectral analysis of the
light-curves and, correspondingly, a spatial mapping of the
magnetospheric gaps (their altitude above the neutron star surface
being possibly related to the shape of their spectra).

We finally note that the timing calibration and tests presented in
this paper pave the way to an effective search for new gamma-ray
pulsars with \emph{AGILE}. The negligible discrepancies among the
radio and the gamma-ray pulsar spin parameters seen in the known
gamma-ray pulsars imply that the direct folding of the \emph{AGILE}
data on new pulsar candidates is a safe and efficient procedure, when
the folding parameters are obtained from radio/X-rays ephemeris having
suitable epoch range of validity.  In fact, gamma-ray period search
trials around radio/X-rays solutions will be unnecessary when
simultaneous ephemeris are available, strongly improving the detection
significance for faint sources.  According to the predicted gamma-ray
pulsar luminosities ($L_\gamma \propto \dot{E}^{1/2}/d^2$,
e.g. \citealt{pellizzoni04agile}), the detection with \emph{AGILE} of
top-ranked Vela-like pulsars is then expected as soon as exposure
levels $\gtrsim$$10^9$ cm$^2$ s ($E>100$ MeV) will be attained for
clean G-class events.

\acknowledgments
We acknowledge D.~A.~Smith, D.~J.~Thompson and G.~Tosti of GLAST Team for
useful discussions on multi-wavelength observations of pulsars and for
their comments on the paper draft.  JPH was supported by NASA 
\emph{XMM-Newton} grants NNX06AH58G and NNX07AU65G. AP and MB received 
financial support from the Italian Minister of Research (MIUR) under national
program PRIN-MIUR 2005. The Parkes radio telescope is part of the
Australia Telescope which is funded by the Commonwealth of Australia
for operation as a National Facility managed by CSIRO.
\emph{XMM-Newton} is an ESA science mission with instruments and 
contributions directly funded by ESA Member States and NASA.

\bibliographystyle{apj}
\bibliography{biblio}

\end{document}